\documentstyle[sprocl]{article}

\input{psfig}

\def\Journal#1#2#3#4{{#1} {\bf #2}, #3 (#4)}

\def\PLB{{\em Phys. Lett.}  B}
\def\PRL{\em Phys. Rev. Lett.}
\def\PRD{{\em Phys. Rev.} D}

\def\be{\begin{equation}}
\def\ee{\end{equation}}
\def\bea{\begin{eqnarray}}
\def\eea{\end{eqnarray}}

\begin{document}

\title{
Prompt $J/\psi$ Polarization 
at the Tevatron}

\author{Jungil Lee\footnote{
Talk given at 5th Workshop on QCD,
3-7 January 2000, Villefranche-sur-Mer, France.
}}

\address{II. Institute of Theoretical Physics,\\
University of Hamburg, 22761 Hamburg, Germany\\
E-mail: jungil@mail.desy.de}

\maketitle\abstracts{
The polarization of prompt $J/\psi$ at the Tevatron is calculated
within the nonrelativistic QCD (NRQCD) factorization framework.
The contribution from radiative decays of P-wave charmonium states decreases,
but does not eliminate,
the transverse polarization at large transverse momentum.
The prediction agrees with measurements
from the CDF collaboration at intermediate values of $p_T$,
but disagrees at the large values of $p_T$ measured.
}

The NRQCD factorization approach to inclusive quarkonium 
production~\cite{B-B-L} 
makes the remarkable prediction that in hadron collisions
$J^{PC} = 1^{--}$ quarkonium states
should be transversely polarized at large $p_T$~\cite{Cho-Wise}.  
In high-energy $p \bar p$ collisions,
the dominant contribution to the charmonium production rate at large $p_T$
comes from gluon {\it fragmentation}~\cite{B-F}.
At leading order in $\alpha_s$,
a $Q \overline{Q}$ pair with small relative momentum
created by the virtual gluon is
in a color-octet $^3S_1$ state 
with the same transverse polarization as the almost on-shell gluon.
Due to the approximate heavy quark spin symmetry of NRQCD,
$1^{--}$ quarkonium states from the pair should have a 
large transverse polarization at sufficiently large $p_T$.
Recent measurements at the Tevatron by the CDF collaboration seem to be in 
contradiction with this prediction~\cite{CDF-psipol}.

A convenient measure of the polarization is 
$\alpha = (T-2L)/(T+2L)$, where $T$ and $L$ are the transverse
and longitudinal polarization fractions in the hadron CM frame.
The variable $\alpha$ 
describes the angular distribution of leptons from the decay
of the $J/\psi$ with respect to the $J/\psi$ momentum.
Beneke and Rothstein studied the dominant fragmentation mechanisms for
producing longitudinally polarized $1^{--}$ states~\cite{Beneke-Rothstein}.
For charmonium production at the Tevatron, 
one should also take into account the {\it fusion} 
contributions from parton processes $i j \to c \bar c + k$
since fragmentation does not yet dominate for most of the $p_T$ region.
The polarization variable $\alpha$
for direct $J/\psi$ and direct $\psi'$ at the Tevatron
have been predicted by Beneke and
Kr\"amer~\cite{Beneke-Kramer} and by Leibovich~\cite{Leibovich}.
They predicted that $\alpha$ should be small for $p_T$ below about 5 GeV,
but then should begin to rise dramatically.
On the contrary, the CDF data shows no sign of transverse polarization 
of direct  $\psi'$
at large $p_T$~\cite{CDF-psipol}.

The CDF collaboration has also measured the
polarization of {\it prompt} $J/\psi$~\cite{CDF-psipol} 
({\it i.e.} $J/\psi$'s that do not come from the decay of $B$ hadrons).
The advantage of this measurement is that the number of $J/\psi$ events
is larger than for $\psi'$ by a factor of about 100.
On the other hand, theoretical predictions of the polarization of 
prompt $J/\psi$ are complicated since the prompt signal includes
$J/\psi$'s that come from decays of the higher charmonium
states $\chi_{c1}$, $\chi_{c2}$, and $\psi'$~\cite{CDF-chi,CDF-chi12}.
The polarization of $J/\psi$ from $\psi'$ not via $\chi_c$ is
straightforward to calculate, since the spin is unchanged by the transition.
The polarization of $J/\psi$ from $\chi_c$ and of
$J/\psi$ from $\psi'$ via $\chi_c$ is more complicated,
because the $\chi_{cJ}$'s are produced in various spin states
and they decay into $J/\psi$ through radiative transitions~\cite{BKL}.
Therefore, it is interesting to investigate the cascade effect in this problem.

The {\it NRQCD factorization formula} for the differential cross section for
the inclusive production of a charmonium state $H$ of momentum $P$
and spin quantum number $\lambda$ has the schematic form
\begin{equation}
d \sigma^{H_\lambda(P)} \;=\;
d \sigma^{c \bar c_n(P)} \;
        \langle O^{H_\lambda(P)}_n \rangle,
\label{sig-fact}
\end{equation}
where the implied sum on $n$ extends over
all the color and angular momentum states of the $c\bar c$ pair.
The $c \bar c$ cross sections $d \sigma^{c \bar c_n}$, which are independent of
$H$, can be calculated using perturbative QCD.
All dependence on the state $H$ is contained within the
nonperturbative NRQCD matrix elements(ME's) 
$\langle O^{H_\lambda(P)}_n \rangle$.
The Lorentz indices, which are suppressed in (\ref{sig-fact}),
are contracted with those of $d \sigma^{c \bar c_n}$
to give a scalar cross section.
The symmetries of NRQCD can be used to reduce
the tensor ME's $\langle O^{H_\lambda(P)}_n \rangle$
to scalar ME's $\langle O^H_n \rangle$ that are
independent of $P$ and $\lambda$.
One may calculate the  cross section for polarized quarkonium
once the relevant scalar ME's are known.

In $p \bar p$ collisions,  the parton processes that dominate the
$c \bar c$ cross section depend on $p_T$.
If $p_T$ is of order $m_c$, the {\it fusion} processes dominate.
These include the parton processes $i j \to c \bar c + k$,
with $i,j=g,q,\bar q$ and $q = u,d,s$.
At $p_T$ much larger than $m_c$,
the parton cross sections are dominated by
{\it fragmentation} processes.
To get a consistent LO evaluation  of polarization parameter $\alpha$
in this region, one should take care.
Since we are interested in the LO prediction of $\alpha$, we  should first
get a LO prediction of longitudinal fraction $L/(T+L)$.
The LO contributions to $L$ are order-$\alpha_s^2$
processes $g\to c\overline{c}+g$, while the only
contribution to $T$ is the order-$\alpha_s$ process  
$g \to c \bar c_8(^3S_1)$ \cite{Beneke-Rothstein}.
Inclusion of the longitudinal part in the fragmentation process is 
crucial in the calculation of $\alpha$ even though it has only a small
contribution to the total cross section.
The fragmentation function is evolved using the
standard homogeneous timelike evolution equation.
Since $c \bar c_8(^3S_1)\to \psi_\lambda(nS)$ is included in both
the fragmentation and the fusion process, we interpolate
between the fusion cross section at low $p_T$ and the fragmentation
cross section at high $p_T$~\cite{BKL,Cho-Leibovich}. 
In all the other $c \bar c\to H$ channels, we use the fusion cross section.

The color-singlet ME's can be determined
phenomenologically from the decay rates for
$\psi(nS)\to \ell^+\ell^-$ and $\chi_{c2}\to\gamma\gamma$~\cite{Maltoni}.
The color-octet ME's are phenomenological
parameters and they are determined from production data at the 
Tevatron~\cite{CDF-psi,CDF-chi}.
In case of $\chi_{c}$,
we have a good agreement with the CDF data on the ratio 
$\chi_{c1}/\chi_{c2}$~\cite{CDF-chi12}.
More details on the analysis methods are explained in 
Refs.~\cite{BKL,Kniehl-Kramer}.
There are numerous theoretical uncertainties in the polarization 
calculation. In our analysis, we allow the variations in
PDF (CTEQ5L and MRST98LO)~\cite{PDF}, 
factorization and fragmentation scales ($\mu_T/2- 2\mu_T$),
charm quark mass ($1.45-1.55$ GeV), respectively,
where $\mu_T = (4 m_c^2 + p_T^2)^{1/2}$.
We also take into account the uncertainties in the ME's,
$\langle O_8(^1S_0) \rangle$ and 
$\langle O_8(^3P_0) \rangle$~\cite{BKL}.

\begin{figure}
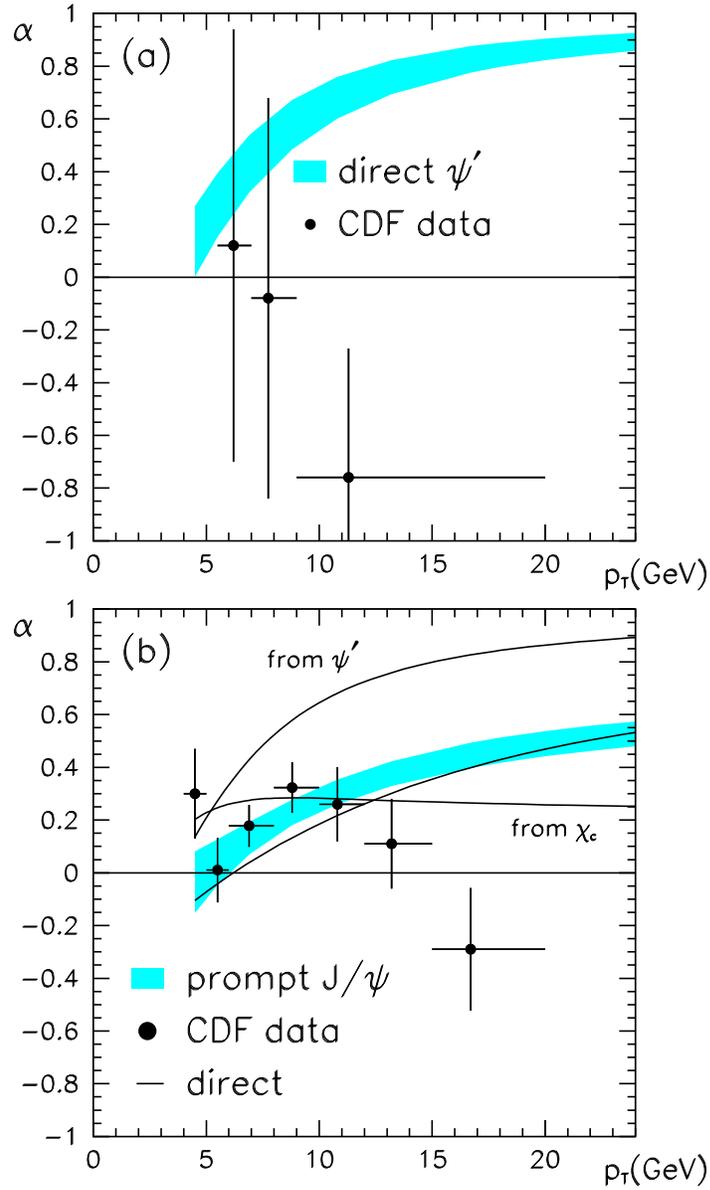

\begin{center}
\begin{tabular}{c}
\psfig{file=lee1a.epsi,height=7.8cm}\\
\psfig{file=lee1b.epsi,height=7.8cm}
\end{tabular}
\end{center}
\caption{
Polarization variable $\alpha$ vs. $p_T$
for (a) direct $\psi'$ and (b) prompt $J/\psi$
compared to CDF data.
}
\end{figure}

Let us compare our results for $\alpha$ with the CDF data.
We present our result in the form of an error band obtained by
combining in quadrature all the theoretical errors described previously.
In Fig.~1(a), we compare our result for direct $\psi'$
as a function of $p_T$ with the CDF data~\cite{CDF-psipol}.
But the error bars in the CDF data are too large to draw any
definitive conclusions.
We next consider the polarization variable $\alpha$ for prompt $J/\psi$.
The method for calculating cascade effect in prompt $J/\psi$ polarization 
is explained in Ref.~\cite{BKL}.
In Fig.~1(b), we compare our result for $\alpha$
as a function of $p_T$ with the CDF data~\cite{CDF-psipol}.
Our result for $\alpha$ is small around $p_T=5$ GeV, but it increases
with $p_T$.
Our result is in good agreement with the CDF measurement
at intermediate values of $p_T$,
but it disagrees 
in the highest $p_T$ bin, where the CDF measurement is consistent with 0.
The solid lines in Fig.~1(b) are the central curves of $\alpha$ for
direct $J/\psi$ and for $J/\psi$ from $\chi_c$. 
The $\alpha$ for direct $J/\psi$ is smaller than that
for direct $\psi'$, because the ME's are different~\cite{BKL}.
In the moderate $p_T$ region,
the contributions from $\psi'$ and from $\chi_c$ add to give an increase
in the transverse polarization of prompt $J/\psi$
compared to direct $J/\psi$.  In the high $p_T$ region,
the contributions from $\psi'$ and $\chi_c$ tend to cancel.

The CDF measurement of the polarization of prompt $J/\psi$
presents a serious challenge to the NRQCD factorization formalism
for inclusive quarkonium production.
The qualitative prediction that $\alpha$ should increase
at large $p_T$ seems inescapable.
However, it is still worthwhile to investigate the NLO effect 
in the fragmentation processes, since the dramatic discrepancy appears
at large $p_T$. 
Recently, Braaten and Lee performed the full NLO calculation of the
color-octet $^3S_1$ gluon fragmentation function 
for polarized heavy quarkonium~\cite{BL}. 
It is not sufficient to get NLO prediction of $\alpha$, but one can first
study the production rate at large $p_T$ in NLO accuracy.

In Run II of the Tevatron, the data sample for $J/\psi$ should be
more than an order of magnitude larger than in Run I,
which will allow both the production rate and the polarization 
to be measured with higher precision and out to larger values of $p_T$.
If the result continues to disagree with the predictions
of the NRQCD factorization approach, it would indicate a serious flaw in our
understanding of inclusive charmonium production.
The predictions of low-order perturbative QCD for the
spin-dependence of $c \bar c$ cross sections could be wrong,
or the use of NRQCD to understand the systematics
of the formation of charmonium from the $c \bar c$ pair could be flawed,
or $m_c$ could simply be too small to apply
the factorization approach to the charmonium system.

\section*{Acknowledgments}
The author thanks Eric Braaten and Bernd A. Kniehl for their enjoyable
collaboration on the subject discussed here. 
This work was supported in part by 
the Alexander von Humboldt Foundation.


\end{document}